\documentclass[traditabstract]{aa} 

\usepackage{graphicx}
\usepackage{natbib}
\usepackage{txfonts}

\newcommand{\lae}{Ly$\alpha$ }
\newcommand{\lya}{Ly$\alpha$ }

\begin{document}

   \title{Inclination Dependence of Lyman-$\alpha$  Properties in a Turbulent Disk Galaxy}

   \author{C. Behrens
          \and
         H. Braun
          }
   \institute{Institut f\"ur Astrophysik, Georg-August Universit\"at G\"ottingen,
              Friedrich-Hund-Platz 1, D-37077 G\"ottingen\\
              \email{cbehren@astro.physik.uni-goettingen.de/hbraun@astro.physik.uni-goettingen.de}}
	
   \date{Accepted in A\&A, October 2014}

  \abstract{We present simulations of Lyman-$\alpha$ radiation transfer in an isolated disk galaxy with a turbulence sub-grid model, multi-phase interstellar medium and detailed star formation modelling. We investigate the influence of inclination on the observed \lya properties for different snapshots. The \lya spectrum, equivalent width distribution and escape fractions vary significantly with the detailed morphology of the disk, leading to variations from one snapshot to another. In particular, we find that supernova-driven cavities near star-forming regions in the simulation can dominate the transmitted \lya fraction, suggesting a variability of LAEs on the timescales of the star formation activity.}

   \maketitle
%

\section{Introduction}
Galaxies detected by their strong Lyman-$\alpha$ (Ly$\alpha$) emission, called \lya emitters (LAEs), have become an important tool in the understanding of cosmology at intermediate to high redshifts ($z > 1.5$), e.g. regarding the large-scale matter distribution \citep{Hill2008,Adams2011} or the epoch of reionization via the connection of LAEs and Lyman-continuum leakers \citep{Behrens2014,Verhamme14,Dijkstra2014}. The physical type(s) of objects classified as LAEs and the connection of the observed line profiles to the physical properties of the emitters are still a question of debate, though progress has been made in recent years \citep[e.g.][]{Ahn03,Verhamme2006,Dijkstra2010}. In particular, scattering on neutral gas shells around HII bubbles and the attenuation of IGM suppressing the blue part of the spectrum can explain many features of the observed spectra, e.g. their asymmetry.

In some recent publications, work has been focused on understanding how anisotropies of LAEs in terms of column densities and velocity fields affect observed \lae properties with respect to the observer's position, mainly for simplified models of LAEs \citep{Zheng13,Laursen2013,Behrens2014,Verhamme14,Gronke2014,Duval2014}. For quantifying the inclination dependence of \lya transmission, more realistic models of LAEs are needed, for example for estimating how large-scale surveys are affected by a possible alignment of galaxies with the large-scale structure in combination with inclination effects \citep{Hirata2009} or other correlations with the large-scale structure \citep{Zheng10,Zheng11,Behrens2013}. While there exists a number of studies of the \lya transport with more realistic LAEs in cosmological contexts \citep[e.g.][]{Tasitsiomi2006,Laursen2009,Yajima2012,Barnes2011,Faucher-Giguere2010} only the work by \cite{Verhamme12} (hereafter: VDB12) investigates the effects of directional dependence on the 
basis of high-resolution, dusty, isolated disk simulations in detail. Their work includes radiative transfer calculations in the resolved interstellar medium (ISM) as well as the transfer of continuum photons 
for the  estimation of the \lya equivalent width (EW). VDB12 find the inclination effect to be strong, leading to variations of the observed EW from -5 to 90 $\AA{}$ for edge-on/face-on observers with considerable scatter. 

In this paper, we extend our previous work on simplified models that were motivated by the existence of optically-thin outflows in observed galaxies to a more realistic setup and present a new study on the radiative transfer of \lya and continuum photons in an isolated disk galaxy with an advanced model of star formation, turbulence and feedback. In particular, we use snapshots of our model galaxy spanning 1 Gyr of its evolution to quantify differences in \lya properties. In the following, we proceed to give a brief summary of the physics used in the simulations presented in \cite{Braun2014} (hereafter: BSN14) in section \ref{sims}. We explain how we post-processed the resulting snapshots from these simulations with our Lyman-$\alpha$ code in section \ref{rt}. In section \ref{res}, we present our results, followed by discussion and conclusions in section \ref{discussion}.
\section{Simulations of Isolated Disk Galaxies}
\label{sims}
We post-process snapshots from a simulation of an isolated disk galaxy that was presented by BSN14 as the 'ref' run. We refer the reader to this paper for the details and briefly summarize the main ingredients of these simulations here. The simulations were performed using the adaptive mesh refinement (AMR) code \texttt{Nyx} \citep{Almgren2013} with an effective resolution of $\sim$ 30 pc. The simulated galaxy resides within a box of 0.5 Mpc size and is initialized as a purely gaseous disk without stars using an adiabatically stable, isothermal setup. Employed physics include
\begin{itemize}
 \item \textit{self gravity} from gas and stars and additionally the gravitational potential of a static NFW-shaped dark matter halo
 \item \textit{gas dynamics} using the piecewise parabolic method for the resolved motions and a subgrid scale model for unresolved turbulence following \cite{Schmidt2011}
 \item \textit{cooling} from gas, metals, and dust
 \item \textit{multiphase ISM} consisting of a diffuse warm, a clumpy cold, and a hot phase for supernova ejecta
 \item \textit{star formation} at a rate depending locally on the inferred molecular fraction and the thermal/turbulent state of the gas
 \item \textit{stellar feedback} in the form of a combination of thermal and turbulent feedback from supernovae (SN) and thermal feedback from Lyman continuum heating both depending on the age of the stars.
\item \textit{metal enrichment} due to SN
\end{itemize}
We stress that gas dynamics, multiphase ISM, star formation and stellar feedback model all couple to the turbulent subgrid model. The simulation covers 2 Gyr of the evolution of the disk. Starting from an initially smooth configuration, the gaseous disk evolves into a flocculent disk with transient spiral features due to the dynamical self-regulation of star formation and stellar feedback. With time a rather smooth and extended stellar disk forms that features several short lived and very few longer-lived stellar clusters. Along with the enrichment with metals, the SN feedback also drives a galactic outflow carrying colder disk material with it that mostly falls back onto the disk. 

We use three different snapshots from the simulation, representing the state of the galaxy 1, 1.5, and 2 Gyr after initialization of the disk. In this regime, the star formation has settled to a self-regulated state. Details about these snapshots are given in table \ref{table1}.

\begin{table*}
\caption{Physical parameters of the used snapshots}             
\label{table1}      
\centering                          
\begin{tabular}{c c c c c c}        
\hline\hline                 
   Age & $M_{g}$($M_{\odot}$)&  $M_{HI}$ ($M_{\odot}$)& $Z$ ($Z_{sol}$) & $M_D$ ($M_{\odot}$)& SFR ($M_{\odot}/yr$) \\    
   1 Gyr & $6.8 \times 10^{9}$ & $4.1 \times 10^9$ & 0.47 & 1.1 $\times 10^7$& 2.8\\      
   1.5 Gyr  & $5.8 \times 10^{9}$ & $3.4 \times 10^9$& 0.56 & 1.08 $\times 10^7$ &2.2 \\
   2 Gyr  & $5.1 \times 10^{9}$ & $ 2.9 \times 10^9$ & 0.58 & 0.96 $\times 10^7$ & 1.7 \\
   
\hline                                   
\end{tabular}
\end{table*}

\section{Lyman-$\alpha$ Transport Post-Processing}
\label{rt}
We use a standard Monte-Carlo approach to post-process the BSN14 simulations. We only summarize the additional physics and the implementation of the radiative transfer here, for a more thorough description we refer the reader to \cite{Dijkstra2006}, \cite{Verhamme2006}, \cite{Behrens2013} among others.

Our code which, first presented in \cite{Behrens2014}, is based on the \texttt{BoxLib}\footnote{https://ccse.lbl.gov/BoxLib/} framework which is also the basis of \texttt{Nyx} \citep{Almgren2013}. This makes it straightforward to post-process the BSN14 simulation data, apart from a few physical prescriptions that we need for the purpose of performing the radiative transfer. We describe these in detail in the following paragraphs. For comparison with VDB12, we closely follow their setup if possible.
\subsection{Ionization State}
The BSN14 simulations do not explicitly trace the ionized fraction of neutral hydrogen. We therefore use the publicly-available code \texttt{Cloudy}\footnote{http://www.nublado.org/ (Version 08.00)} to produce tables that yield the neutral fraction as a function of thermal energy, metallicity and total density in collisional ionization equilibrium \citep{Ferland1998}. 
\subsection{Emissivity}
In order to run the Monte-Carlo simulation, we need to prescribe an initial spatial and spectral distribution of tracer photons on the AMR grid. The BSN14 model already contains a field for the emissivity of Lyman continuum (LyC) photons from young stars. These trace the number density of Lyman-$\alpha$ photons, since ionization of neutral hydrogen by LyC followed by case-B recombination yields Lyman-$\alpha$ radiation. We assume here that this conversion happens locally, resulting in a direct proportionality between LyC photon density and Lyman-$\alpha$ photon density. As input spectrum, we use a Gaussian with a width of 10 km/s, centered around zero in the restframe of the given gas cell.

Additionally, we follow the radiative transfer of continuum photons near the \lya line to calculate the EW distribution. Continuum photons are emitted with the same spatial distribution as Lyman-$\alpha$ photons, but with a flat spectrum extending out to $\pm 2\times10^4$ km/s around the \lya line center. The total number of \lya/continuum photons launched per snapshot is about $1.8 \times 10^7$. 
\subsection{Lyman-$\alpha$ Equivalent Width}
To infer EWs from our radiative transfer simulations, we need to determine the ratio of emitted line and continuum flux which is equivalent to setting the intrinsic EW. This translates to a condition for the constant luminosity of continuum photons $L_c$ per wavelength bin $\Delta \lambda$ and the total intrinsic Lyman-$\alpha$ luminosity $L_\alpha$,
\begin{equation}
 \frac{\Delta L_c}{\Delta \lambda} = L_\alpha/E
\end{equation}
where $E$ is the desired intrinsic EW. In terms of the implementation, this condition fixes the ratio of continuum- and Lyman-$\alpha$ photons. We set the intrinsic EW to 200 \AA{}. We note that we do not derive this value from the star formation rate and the stellar mass.

\subsection{Dust Distribution}
Since the BSN14 simulations track the metallicity $Z = \rho_m/\rho$ where $\rho$ is the total density and $\rho_m$ is the metal density, but not the dust grain number density $n_D$, we have to implement a model for the distribution of dust. We slightly modify the model used by VDB12 (their eq. 3). The fraction of metals that settles in dust grains is assumed to be proportional to the neutral fraction of hydrogen $f_{N}$ and to the dust-to-metal ratio $R_{\mathrm{dust/metal}}$ which is assumed to be 0.3 based on an empirical value for our own Galaxy \citep{Inoue2003}:
\begin{equation}
 n_D = f_N R_{\mathrm{dust/metal}} Z \rho/m_{grain}
\end{equation}
Since the transition between the fully ionized and the fully neutral state is sharp, this model yields effectively similar results to the one used by VDB12. Employing their model would change the dust content only by a few percent.
VDB12 use a value of $3 \times 10^{-17}$ g for the typical grain mass. Employing this value of the dust grain mass renders our galaxy a complete \lya absorber (at least for the 1 Gyr snapshot) with EWs consistently below zero. This is plausible since our galaxy is about a factor of 10 more massive than the one in VDB12. Since we are interested in the detailed inclination effect here and have no access to a less massive galaxy run, we scaled the total dust grain number density in our galaxy by a factor of about 25, i.e. reducing the optical depth due to dust by this factor. This was implemented by simply scaling the dust grain mass to $m_{grain} = 8 \times 10^{-16}$ g without changing the cross section. We call this model the 'fiducial' one, but we comment on the results for the runs with the VDB12 model that we will call the 'realistic' runs.

 For the interaction cross section and reemission phase function, we follow \cite{Schaerer2011}, using similar parameters, i.e. an albedo $Q=0.5$, and a Greenstein phase function parameter $G = 0.7$. 
\subsection{Influence of the Turbulent Subgrid Scale Energy}
The Doppler parameter $b$ governs the width of the associated Voigt profile. The turbulent contribution to it is introduced as 
\begin{equation}
 b = \sqrt{v_{th}^2+v_{turb}^2}\,
\end{equation}
where $v_{th}$ is the typical (RMS) thermal velocity in the gas, while $v_{turb}$ is the turbulent velocity component in the cell. While VDB12 assume a constant turbulent velocity here, we derive the typical (RMS) turbulent velocity of each cell directly from its turbulent energy $E_t$, 
\begin{equation}
 v_{turb} = \sqrt{2E_t}
\end{equation}
\subsection{Morphology}

\begin{figure*}
   \centering
   \includegraphics[width=0.6\linewidth]{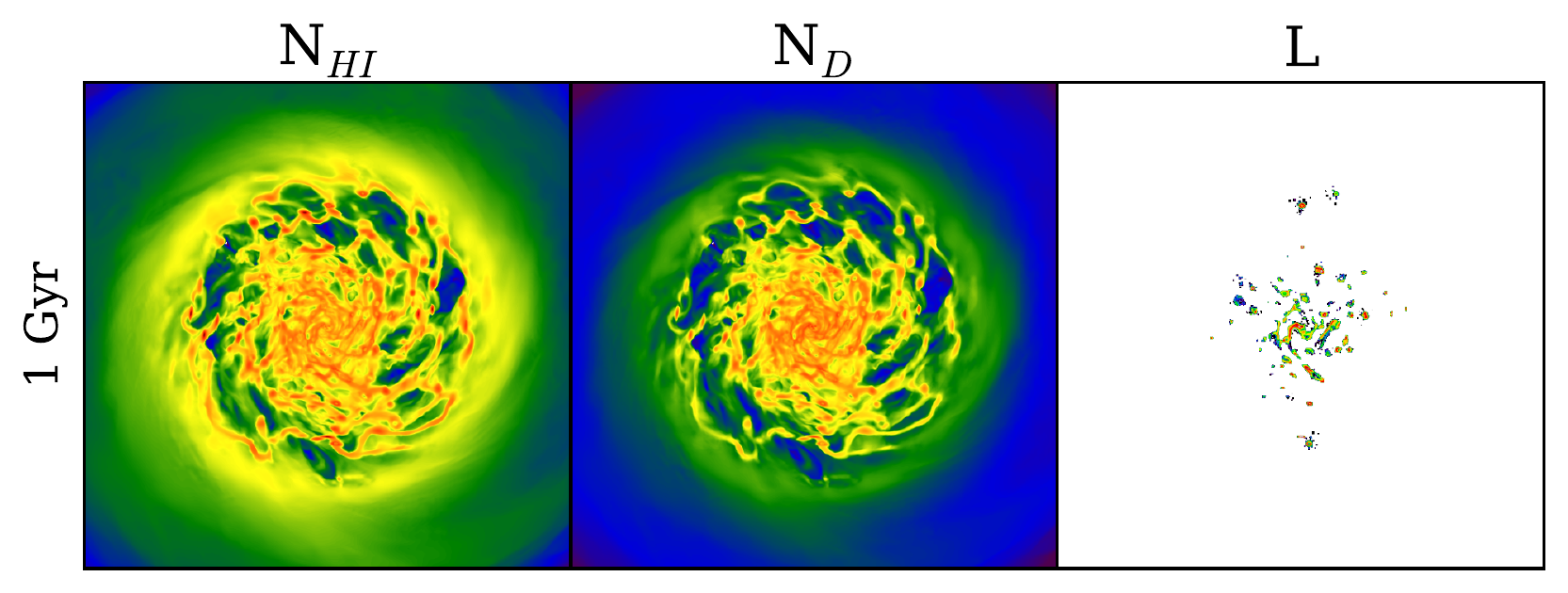}
   \includegraphics[width=0.6\linewidth]{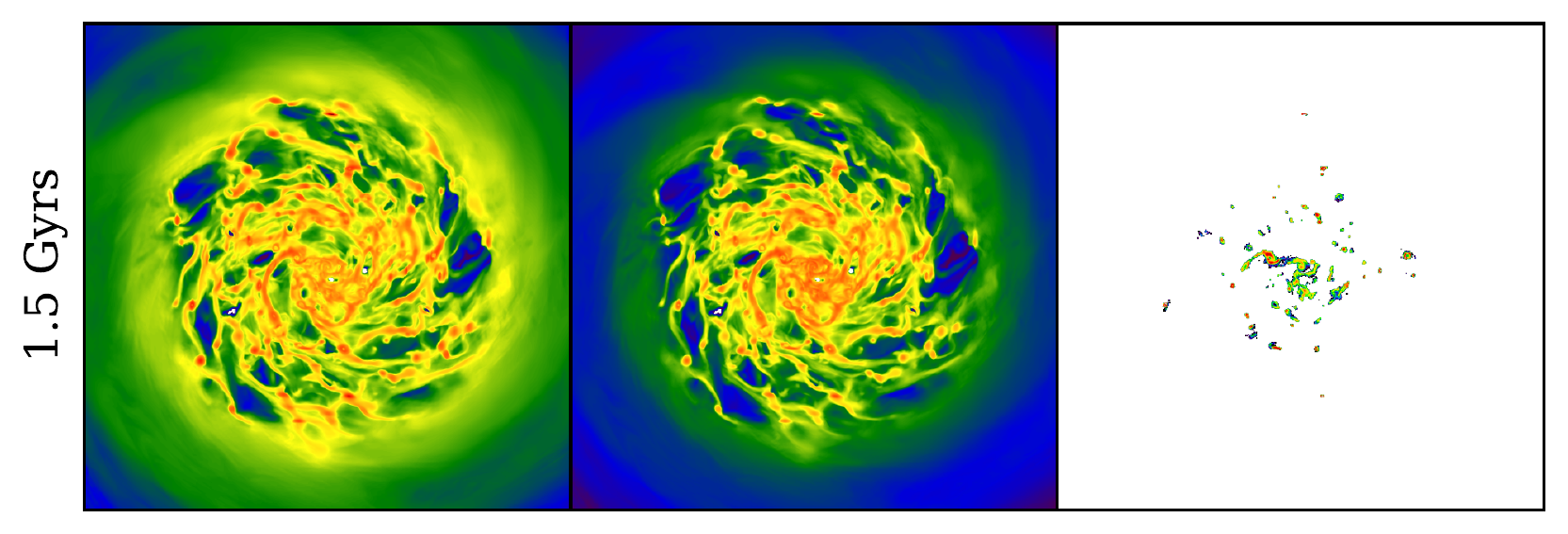}
   \includegraphics[width=0.6\linewidth]{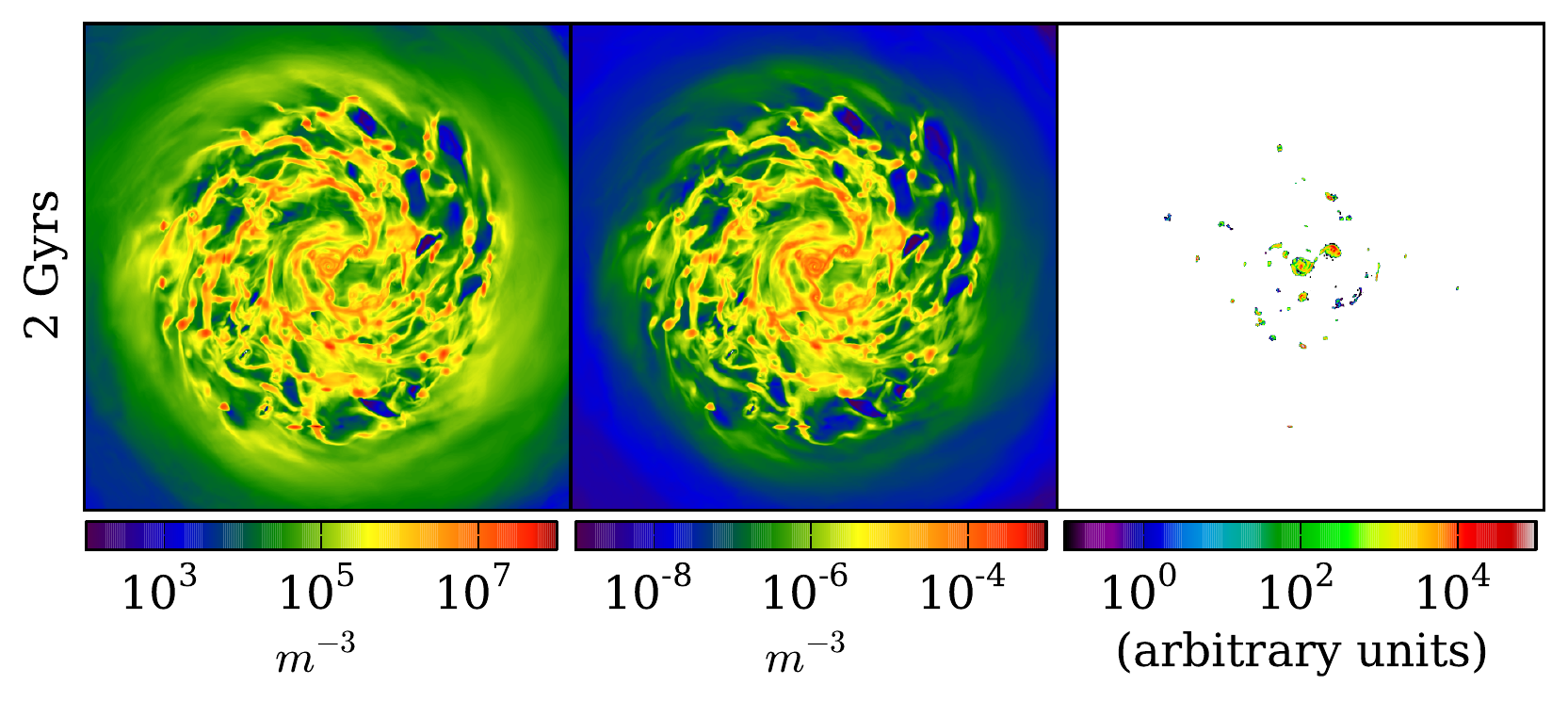}
      \caption{Projected neutral hydrogen density/dust density/emissivity in the disk (left/center/right) for the three snapshots at 1/1.5/2 Gyr (top/center/bottom). Each plot shows the central 40 kpc of the disk.}
         \label{fig:disk1}
\end{figure*}

In Fig. \ref{fig:disk1}, we show the projected HI density, the projected dust density, and the projected distribution of emission resulting from the prescriptions outlined above. The 1 Gyr snapshot (upper plots) shows an unordered density structure, a dense network of individual bubbles and walls. In comparison, the 2 Gyr snapshot (bottom plots) shows a concentrated profile with larger underdense and overdense regions and hints of spiral structures. While the 1 Gyr snapshot shows extended emission in smaller clumps (top right), the emission in the 2 Gyr snapshot (bottom right) is more concentrated to a few clumps near the center. The 1.5 Gyr snapshot lies in between. As expected, the dust density approximately follows the neutral hydrogen density in all cases. As implied by the data in table \ref{table1}, the dust content does not vary significantly between the snapshots. This implies that at least at this stage, the accumulation of metals does not lead to an increase in dust mass due to hots winds carrying 
metals away from the cool disk where dust grains could form.

\section{Results}
\label{res}

\subsection{The Fiducial Simulations}
\begin{figure}
   \centering
   \includegraphics[width=\linewidth]{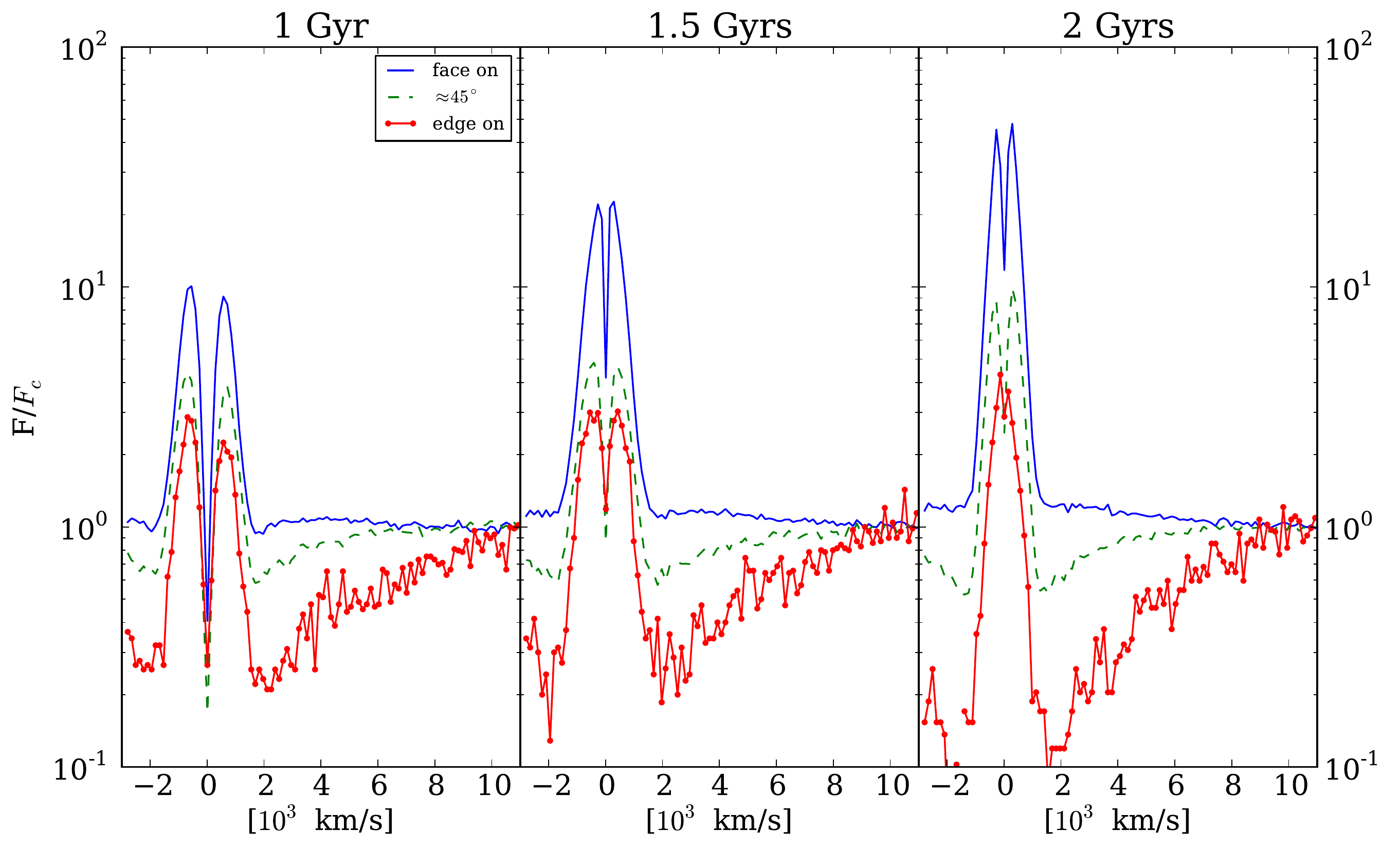}
   
   \caption{Emerging spectra for all three snapshots (left to right: 1/1.5/2 Gyr). We normalize the spectral flux to the continuum far away from the line and show it for three different inclinations: Face-on (solid line), edge-on (dotted line) and at about $45^\circ$ (dashed line).}\label{fig:4}
\end{figure}
The spectra that emerge using the fiducial prescriptions outlined above for the three snapshots (left/center/right panel for 1/1.5/2 Gyr) are shown in Fig. \ref{fig:4}. Each panel features the spectrum obtained in face-on/edge-on direction (solid line/dots) and for $\approx 45^\circ$ with respect to the disk (dashed line). All spectra resemble the typical double-peak spectrum with slightly asymmetric peaks. The double peak, however, is barely visible in the 2 Gyr data. Comparing the three snapshots, we find that the normalized line flux is largest in the 2 Gyr run viewed face-on. The trough at line center is very prominent in the 1 Gyr snapshot and gradually decreases in depth and width in the two other snapshots. The position of the strongest peak changes; while it is the red peak for 1 Gyr and 1.5 Gyr and the edge-on spectrum for 2 Gyr, the face-on spectrum for 1.5 Gyr, the face-on and the $\approx 45^\circ$ spectrum at 2 Gyr show a slightly higher blue peak. As VDB12 point out, the red peak 
might be enhanced by scattering in outflowing material \citep[also see][]{Verhamme2006}. Inflowing material would produce a blue peak. We believe the peaks of the simulated spectra to be related to the fact that both inflowing and outflowing gas exist at the same time. Outflowing gas is driven outwards by feedback, and may fall back onto the disk later. They vary in strength and direction according to the recent star formation history and the entailing stellar feedback in the disk.

\begin{figure}
   \centering
   \includegraphics[width=\linewidth]{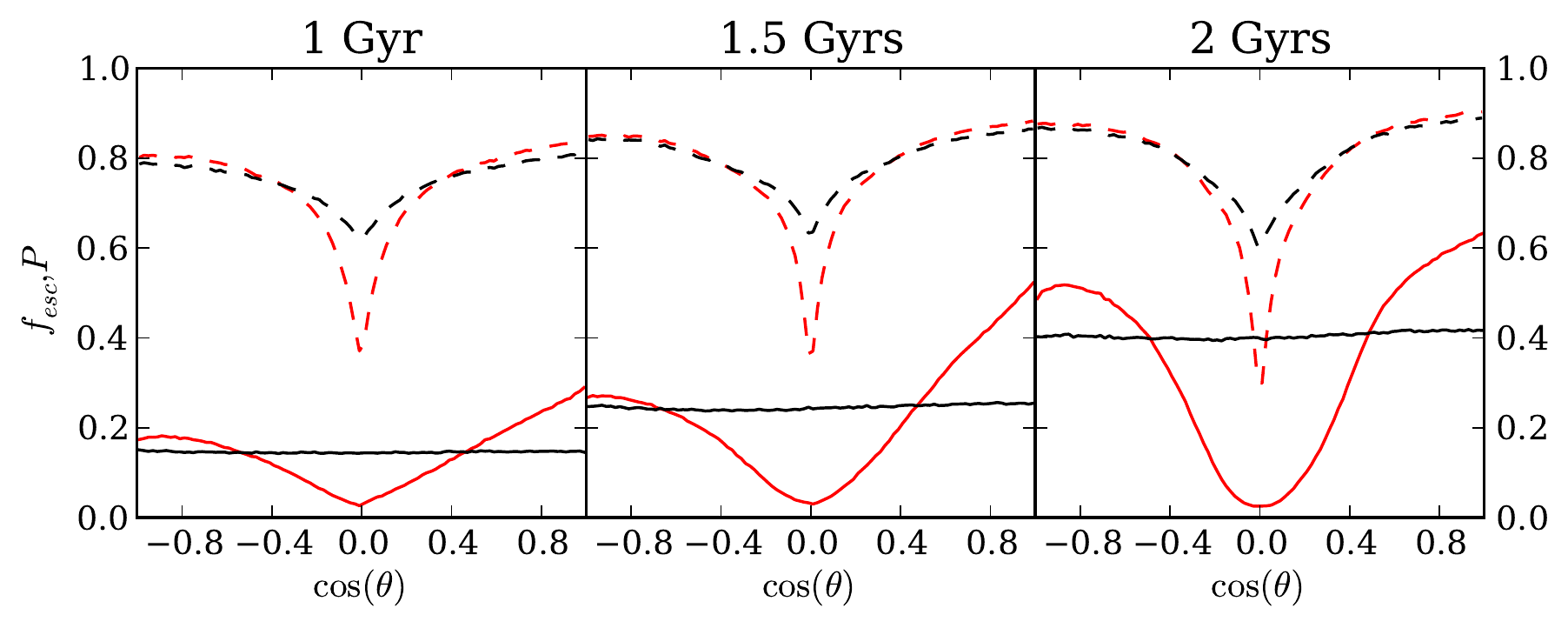}
   
   \caption{Escape fraction as a function of inclination at escape in gray/red for continuum (dashed) and \lya (solid). In black we show the probability of transmission given that the photon was initially emitted with inclination $\cos\theta$ again for continuum (dashed) and \lya (solid). Data is shown for the 1 Gyr/1.5 Gyr/2 Gyr snapshot from left to right.}\label{fig:1}
\end{figure}

Fig. \ref{fig:1} shows the distribution of escape directions (gray/red) and the probability of a photon escaping with a certain initial inclination (black) as a function of $\cos \theta$, where -1 and 1 correspond to face-on from below (along the $-z$-axis) and above the disk (along the $+z$-axis). Edge-on corresponds to a value of $\cos \theta =$ 0. We show the data for both continuum photons (dashed) and \lya photons (solid) for each of the three snapshots (1/1.5/2 Gyr from left to right). For \lya photons, the probability to escape is independent of the initial direction of the photon, as also found by VDB12 (their Fig. 8). Similar to their results, \lya photons show in general a strong tendency to escape face-on, but we observe that the slope is different for negative/positive inclinations. For the continuum photon, both a weak dependence of the escape fraction and the escape direction on inclination is found, similar to VDB12. The amplitude of the inclination effect is different, and the exact shape and 
steepness of 
the dependency differs. For example, for the 1 Gyr snapshot, the escape fraction for \lya photons is a rather shallow and nearly linear function, but the total range of escape fractions spans about a factor of 10, which is comparable to VDB12. At 2 Gyr, the variation is about a factor of 60, and the distribution is also much steeper. 

The \lya mean escape fraction varies by a factor of $\approx 3$ across 
the snapshots, while the continuum escape fraction approximately remains constant: The mean escape fractions of \lya photons are 14/24/41\% at 1/1.5/2 Gyr for the \lya photons, while 75/79/81\% of the continuum photons escape, respectively. The \lya escape fractions are broadly consistent with former studies at lower resolution \citep[e.g][]{Laursen2009}, but in some tension with VDB12: they infer 22\% for continuum and 5\% for \lya photons (for their simulation G2). This difference is readily explained by the fact that we scaled the dust content down to study the inclination effect. In fact, in the realistic case, we obtain lower escape fractions of 0.2-15\% consistent with our galaxy being more massive.

\begin{figure}
   \centering
   \includegraphics[width=\linewidth]{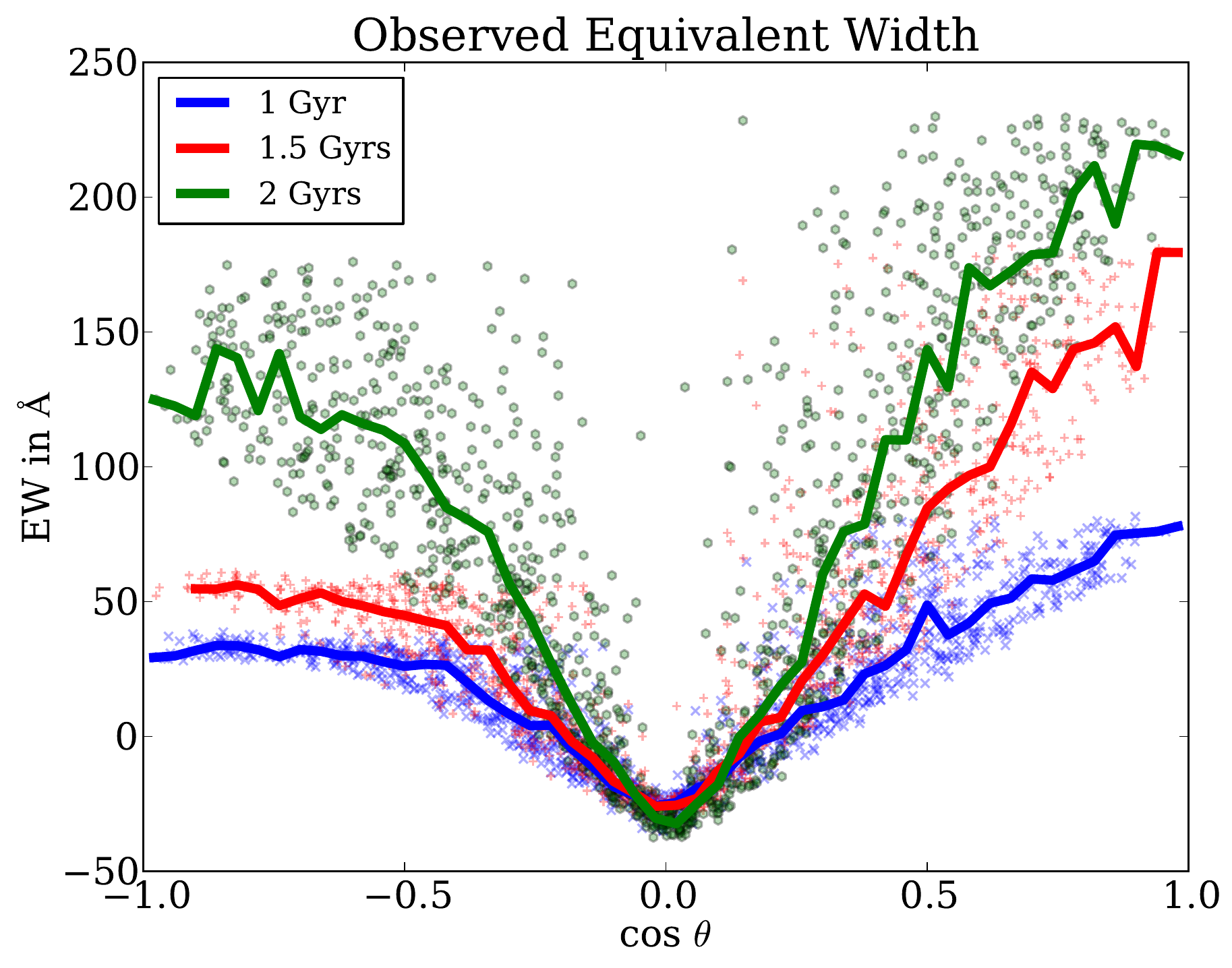}
   
   \caption{\lya EW as it would be observed along lines of sight randomly chosen as a function of $\cos \theta$. The lines show the binned median.}\label{fig:2}
\end{figure}

Fig. \ref{fig:2} shows the EW distribution as a function of $\cos \theta$ where each data point corresponds to a randomly chosen observation direction. The EW varies drastically from one line of sight to another and shows a general trend to be larger at larger absolute inclination as expected. Additionally, it varies systematically from snapshot to snapshot: while the maximum EW is around 70 $\AA{}$ for the 1 Gyr snapshot, it peaks around 230 $\AA{}$ for the 2 Gyr run. The asymmetry between observers above and below the disk is also present here. While for the 1/1.5 Gyr run the left side of the plot, corresponding to an observer situated below the disk, is significantly suppressed, this suppression seems to be reduced in the 2 Gyr run. For each observer, we collected all photons within $\pm$ 10$^\circ$ of the line of sight to have sufficient statistics. We caution the reader that the exact minimum/maximum values and the strength of fluctuations changes depending on this angular range; using $\pm$ 20$^\circ$ 
instead of $\pm$ 10$^\circ$ for example would shift the minimum (maximum) value to about 0 (200) $\AA{}$ for the 2 Gyr snapshots. Comparing with the results of VDB12 (their Fig. 10), similar trends are seen, but both the range of values and the shape of the distribution are different. The scatter in our plots is generally larger, especially for the 1.5 and 2 Gyr snapshot. While it is natural to have different EW distributions in our work and VDB12 since both mean escape fraction and escape fraction as a function of inclination are different, some of the differences might also come from a different choice of the angular range for each observer. A large angular range effectively averages out local fluctuations like cavities with small solid angles.

\begin{figure*}
         \centering
   \includegraphics[width=0.6\linewidth]{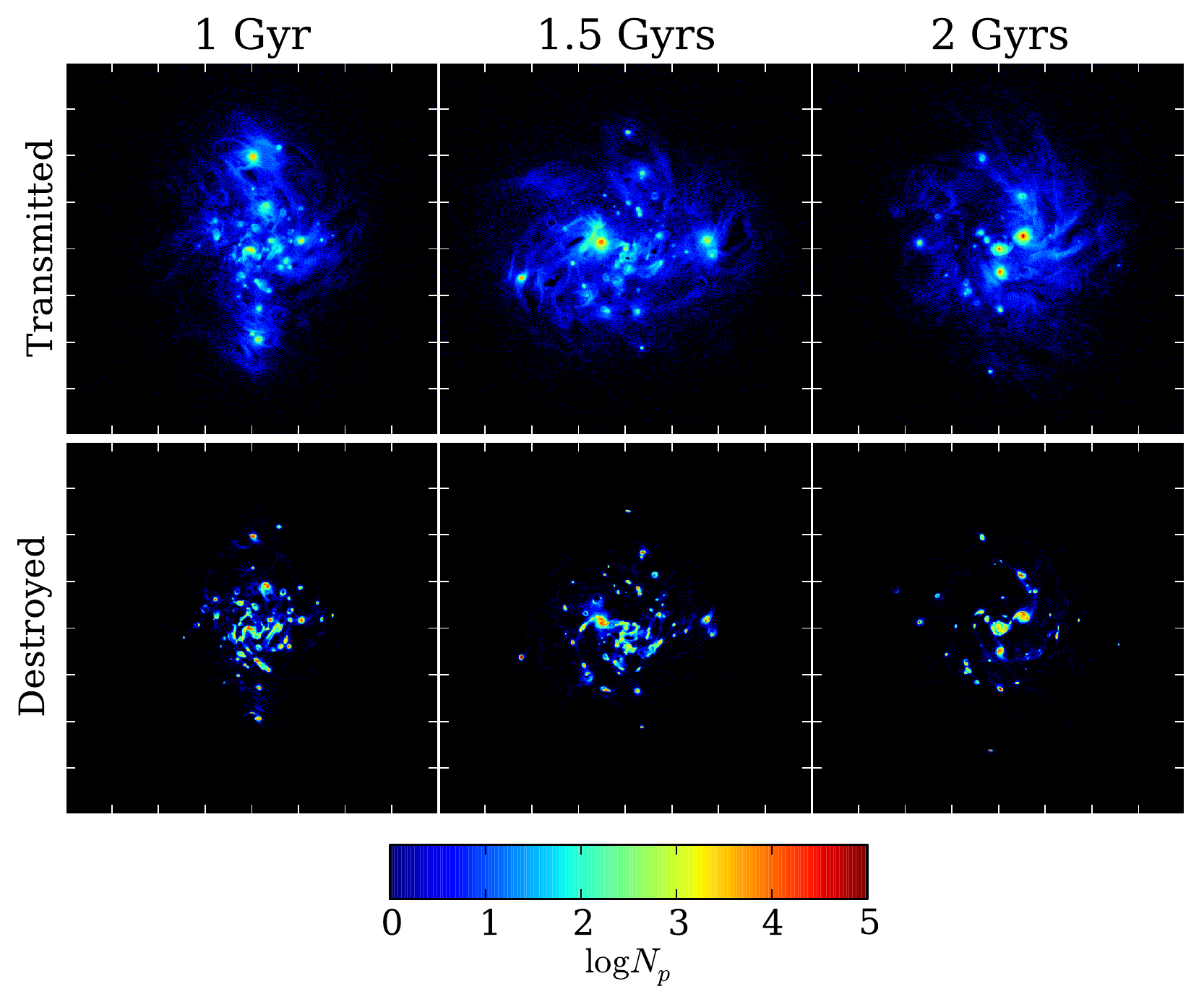}
      \caption{Spatial distribution of escaping photons (top) and absorbed photons (bottom) for the snapshot at 1/1.5/2 Gyr (left/center/right). For the transmitted photons, the distribution is given by the points in the simulation box where photons scattered just before it left the box. For the destroyed \lya photons, we binned the positions at which photons were absorbed. Each plot shows the central 40 kpc of the disk.}
         \label{fig:3}
   \end{figure*}

In Fig. \ref{fig:3}, we show the spatial distribution of both transmitted \lya photons (top) and the photons that were destroyed by dust (bottom plots). The plots show the binned distributions of the spots of last interaction with the gas, i.e. the spots of last scattering (before escape) for the transmitted photons and the locations of absorption for the destroyed photons. For all snapshots, transmission is diffuse except for few clumps that exhibit larger transmission. Comparing the distribution of the transmitted photons with the intrinsic emissivity in Fig. \ref{fig:disk1} (right), we see that most of the emitting clumps do not show transmission. Most of the photons launched in these places are destroyed by dust, which is clearly visible in the distribution of the absorbed photons. The distribution of the destroyed photons in Fig. \ref{fig:3} resembles the emissivity map in Fig. \ref{fig:disk1}, indicating that many of the emitted photons are destroyed locally. This is expected, since the emitting clumps 
are typically very dense.

Comparing the spatial distribution of the transmitted photons among the three snapshots, the transmission is more spatially concentrated in the 1.5 and 2 Gyr snapshots which is partly explained by the fact that the emissivity is more concentrated in these snapshots. Especially for the 2 Gyr snapshot, we see that most of the \lya photons escape from three central clumps, e.g. the largest of those clumps contributes about half of all transmitted \lya photons.

To conclude, we state that there is a distinct top-bottom asymmetry in all snapshots, and that the snapshots differ drastically in terms of \lya properties. In the following, we investigate these results further with focus on the differences between the 1 Gyr and 2 Gyr snapshot. 
\subsubsection{Differences Between the Snapshots}
\label{sec_diff}
From table \ref{table1}, it is clear that the total neutral gas mass decreases with time due to ongoing star formation. Since a smaller neutral gas mass implies a lower optical depth, this partly explains the differences between the subsequent snapshots. One can check this by artificially decreasing the neutral gas mass in e.g. the 1 Gyr snapshot. Reducing the gas mass by a factor of 2 in the 1 Gyr snapshot enhances the EW (and escape fraction) by about 40\% (see section \ref{sec_vary} below), but this is not enough to explain the boost observed between the 1 and 2 Gyr snapshot. This result already gives a hint that the driving factor here is not only the gas mass, but also the detailed morphology of the disk. To understand this local morphology, we analyzed the data of emitted and transmitted photons by detecting large emission spots and correlating them with clumps of escaping \lya photons. Clumps are detected by binning the data onto a grid, applying a 
luminosity threshold and adding up all bins that are spatially connected. This yields a set of emission clumps that have counterparts in transmission and another set of emission spots that lack a counterpart. 

\begin{figure}
   \centering
   \includegraphics[width=\linewidth]{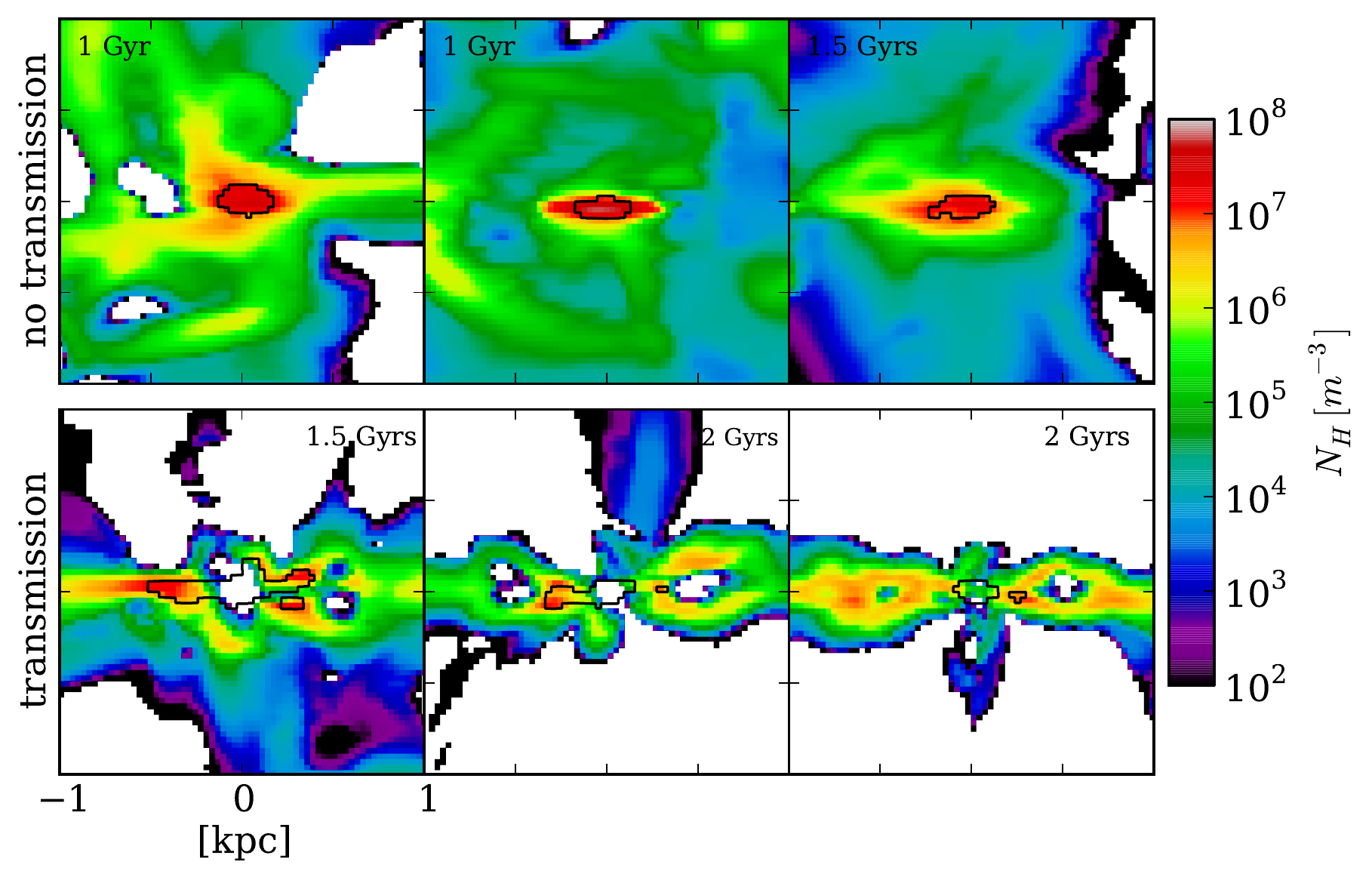}
   
   \caption{Examples for density slices through the disk at the location of various emission spots. The disk lies along the $x$-axis in the plots. Top row: examples for spots that do not exhibit significant transmission. Bottom row: spots that show transmission. The black contours indicate the approximate center of emission in each slice.}\label{fig:slices}
\end{figure}

The high-transmission spots are located in the vicinity of low-density regions due to a highly disturbed environment, or are close to low-density cavities connecting the inner, dense part of the disk with the ionized region around the disk. A few examples of slices of the density distribution around the centers of emission within the disk are shown in Fig. \ref{fig:slices}. The origin of the cavities is the star formation feedback. As turbulent and thermal energy is released into the disk by SN, high-temperature, low-density bubbles form and expand until they drain hot gas and pressure through leakages into their surroundings above and below the galactic disk. For example, the environment of the largest emission clump in the 2 Gyr snapshot (see Fig. \ref{fig:slices}, bottom center) shows a bubble in the center, connected to the outer parts of the disk at the bottom. The column density for direct escape to the top is also low for some lines of sight. Emission spots that are located in a cavity usually have 
counterpart in the transmission, while those that lack a counterpart in the transmitted \lya are typically located in, or surrounded by, dense regions in which the \lya photons scatter until they are eventually absorbed. 

The discrimination of emission spots between those with and without counterparts in transmission is also a distinction between different evolutionary stages of star-forming regions and the environments. In the early stages, the LyC-emitting stars reside in a dense environment. Later, as the first massive stars die in SN explosions a bubble or cavity begins to form which at the same time stalls star formation locally. As the LyC-emitters are short-lived, large cavities are frequently found hosting only negligible \lya emission.

The number of large cavities associated with strong emission spots is lower in the 1 Gyr snapshot compared to that of the 1.5 or 2 Gyr snapshots. This leads, in concert with the already mentioned higher neutral gas content, to the observed lower escape fraction and the lower EW in the 1 Gyr run. The strongest emission spots correlated with cavities found in the late snapshots arise from massive stellar clusters that locally dominate gravity and, as a consequence, host intermittent violent star formation. Those stellar clusters are assembled via mergers of marginally bound clusters, which are the remnants of star forming regions in the low-metallicity regime of star formation, as described by BSN14. On the one hand, the \lya emitting phase in small clusters ends roughly around the time at which the development of a cavity starts, such that the \lya radiation is most probable to be obscured by the clump hosting the star formation originally. \lya radiation from small clusters contributes to the transmitted 
flux only for the short time between the formation of a bubble and the termination of \lya emission after star formation stops (and the formed massive stars die), since the small clusters are created during a single event of star formation without any subsequent star formation. On the other hand, feedback launched at the location of prominent emission spots by stars that were formed during previous duty cycles of a large cluster already prepared a cavity reaching far into the disk's surroundings. This preexisting cavity serves as an escape path for radiation emitted by later generations of stars, such that this kind of emission spot is prominent in the transmission. This explanation implies that for both small and massive clusters, the contribution to \lya transmission varies on the timescale of the star formation activity ($\sim$ few 10 Myr). For massive clusters, this is identical to the timescale of the duty cycle, for small clusters, it is the lifetime of the star-forming region.

\begin{figure}
   \centering
   \includegraphics[width=\linewidth]{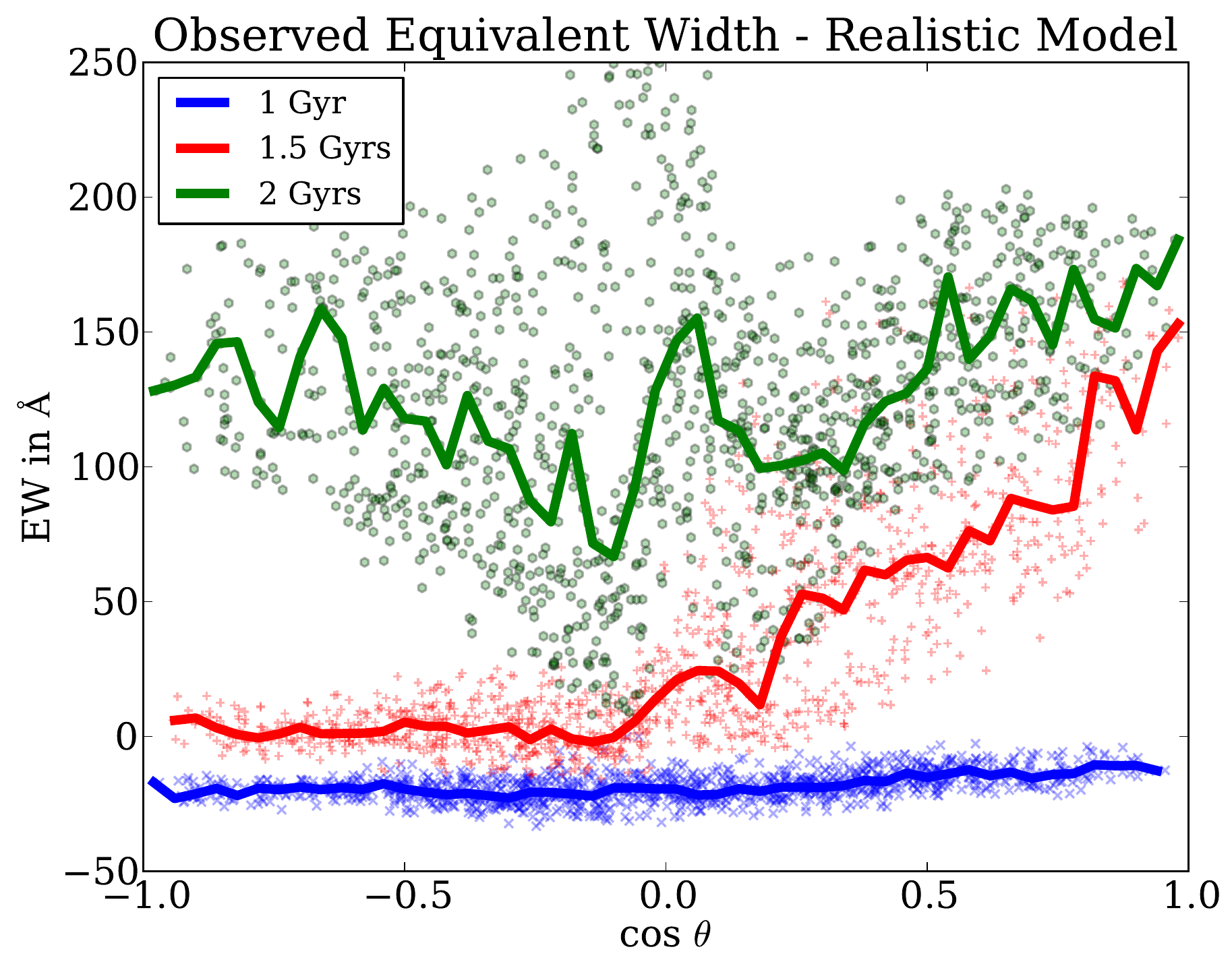}
   
   \caption{Same as fig. \ref{fig:2}, but for the dust prescription used in VDB12, i.e. without the scaling of the dust content we invoked for the fiducial runs.}\label{fig:2_vdb}
\end{figure}

\begin{figure}
   \centering
   \includegraphics[width=\linewidth]{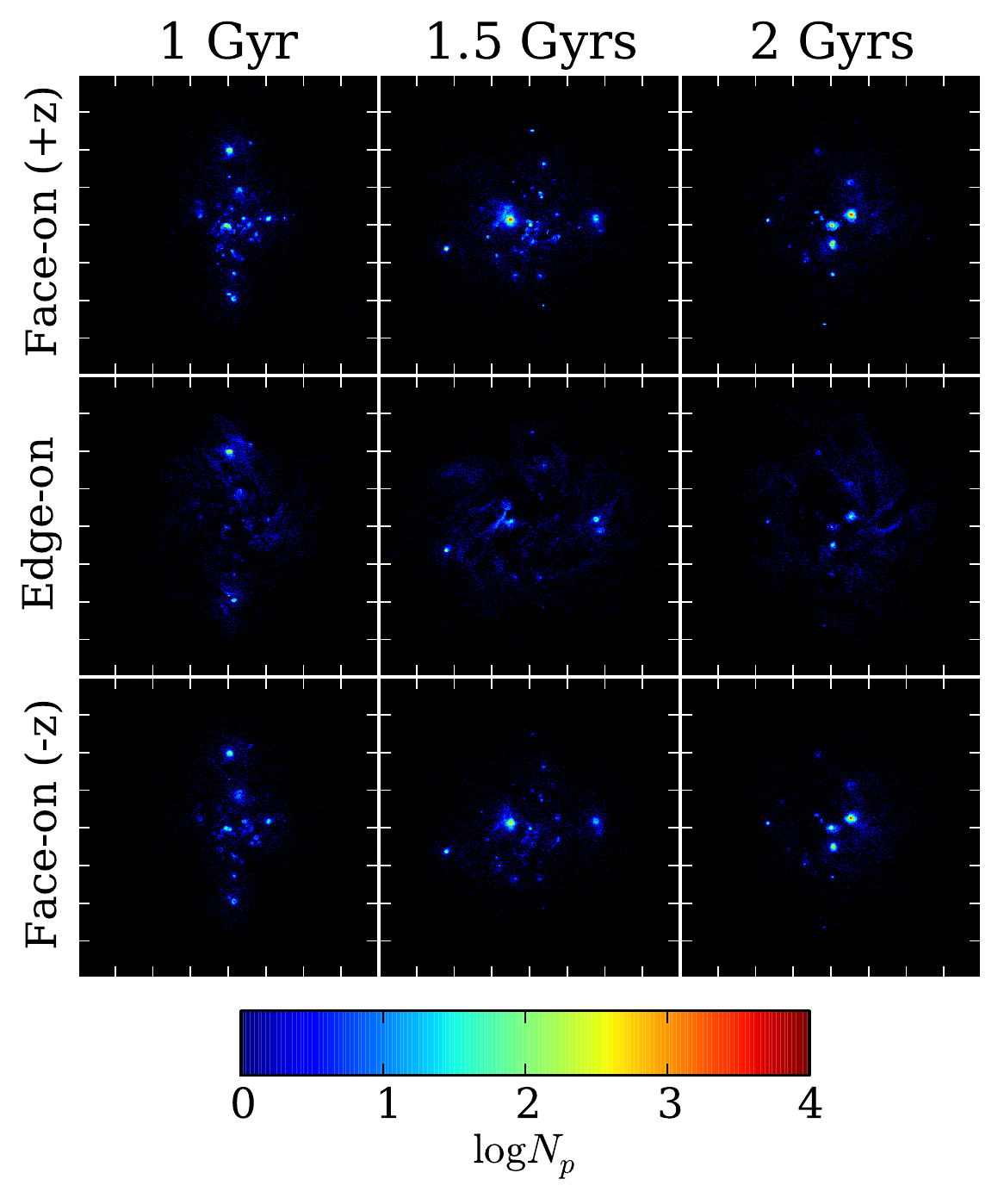}
   \caption{In this plot we show the spatial distribution of transmitted \lya photons for the snapshots at 1/1.5/2 Gyr (left to right). In contrast to Fig. \ref{fig:3}, we show only the photons that escaped face-on towards the positive $z$-axis (top), edge-on (middle), and face-on towards the negative $z$-axis (bottom). Photons within a range of $\pm$10$^\circ$ of the principal direction were considered. Shown are the central 40 kpc of the disk.}\label{fig:3_faceon_edgeon}
\end{figure}

\subsubsection{The Top-Bottom Asymmetry}
The Top-Bottom asymmetry can be readily explained on the basis of the last section. Most of the radiation leaking from the emission spots (see Fig. \ref{fig:slices} for examples) escapes towards the top of the simulation box. 
The excess of the EW at negative $\cos (\theta)$ in the 2 Gyr run compared to the younger snapshots (see Fig. \ref{fig:2}) is caused mostly by the most active emission spot in the 2 Gyr run (see bottom center panel of Fig. \ref{fig:slices}) that transmits equally to both sides ($\pm z$ direction). This is not the case e.g. for the most prominent emission spot in the 1.5 Gyr run (bottom left panel in Fig. \ref{fig:slices}), which contributes about 40\% to the total \lya transmission. To further illustrate this asymmetric behavior, we plot again the distribution of transmitted \lya photons, this time for photons escaping within $\pm$10$^\circ$ to the top/bottom in Fig. \ref{fig:3_faceon_edgeon} (top/bottom panels) for the snapshots (1/1.5/2 Gyr at the left/right). For 
comparison, plots for the photons escaping edge-on (middle row of Fig. \ref{fig:3_faceon_edgeon}) are also shown. Many of the strongly transmitting spots preferentially transmit to the top.

It is plausible that the ongoing feedback processes, namely SN bubbles blowing up, slightly prefer a particular direction even on longer timescales. The feedback-driven bubbles follow the path of least resistance. This path through the surroundings of the disk is likely to be a fossil remnant of the path of another, more or less recent outflow (see Section~\ref{sec_diff}). So if the outflows launched from a particular massive stellar cluster had a top-bottom asymmetry initially due to massive obstacles in one direction, this asymmetry may persist for a few dynamical timescales of the disk, i.e. a few 100 Myr.

\subsection{The Realistic Simulations}
Our results also apply if we do not scale the dust content as we have done for the fiducial runs. In contrast, cavitities become even more effective in boosting escape of \lya photons in this case. We show the EW distribution for the 'realistic' case in \ref{fig:2_vdb}. As has been mentioned before, the 1 Gyr run shows \lya absorption in this case, with a \lya escape fraction of 0.2\% on average. The 1.5/2 Gyr runs, on the other hand, show a very distinct dependence of EW on orientation. Only the cavities identified in the fiducial runs do significantly transmit \lya and increase the average \lya escape fraction to 4 (15)\% in the 1.5 (2) Gyr run. The 1.5 Gyr snapshot therefore shows an enhanced asymmetry related to the orientation of the main transmitting cavity. Similar to the fiducial run, the 2 Gyr run shows transmission both to the top and to the bottom. The strong fluctuations in EWs measured edge-on are an artifact coming from the fact that we do not apply a flux limit: The escape fraction 
for both \lya and continuum photons is below 0.5\% in this region. We conclude that the significance of cavities for the escape of \lya photons is a robust feature.
\subsection{Varying Parameters}
\label{sec_vary}
\begin{figure}
   \centering
   \includegraphics[width=\linewidth]{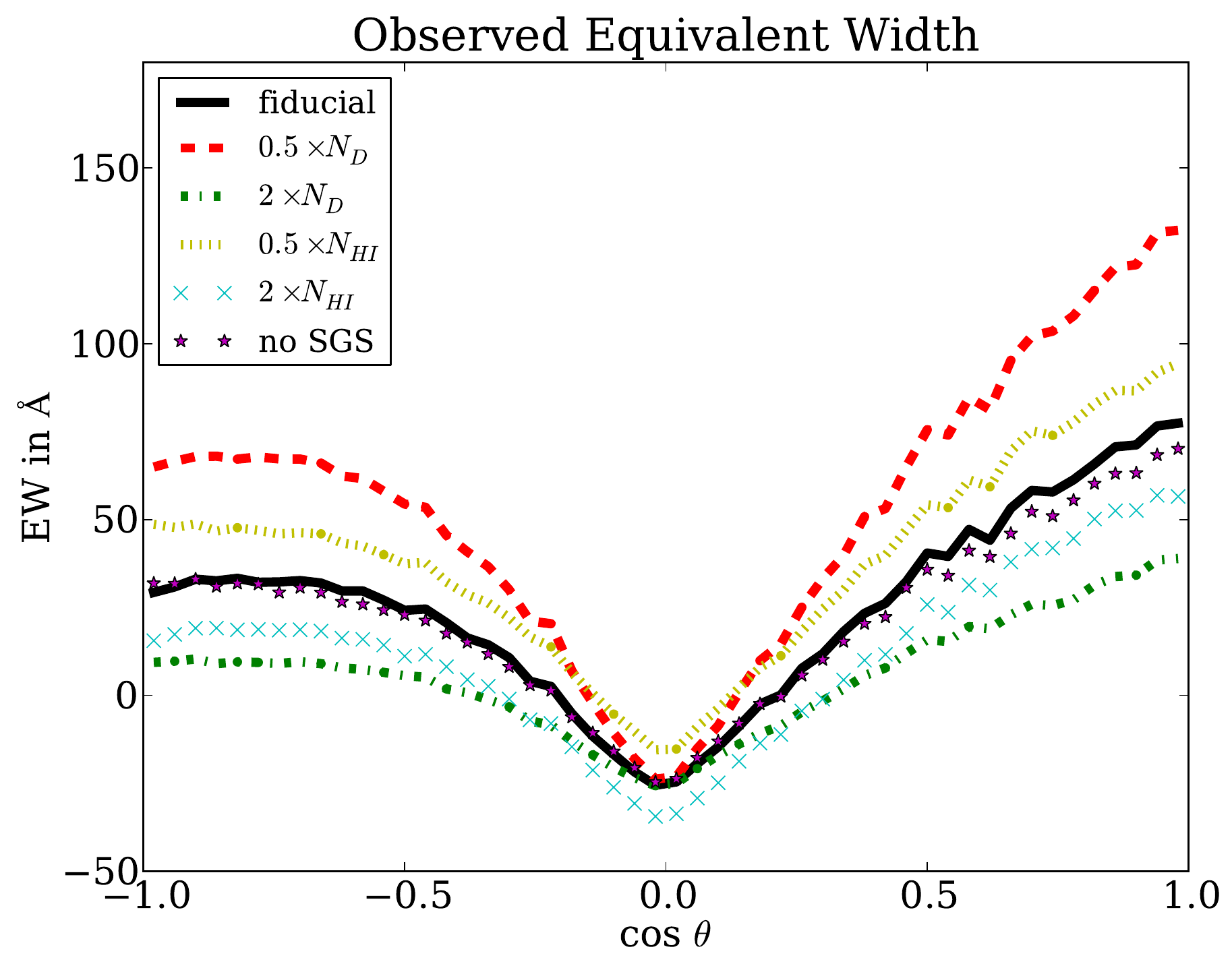}
   
   \caption{Same as Fig. \ref{fig:2}, but for different parameter variations of the 1 Gyr snapshot. We plot the median lines here only.}\label{fig:2_variants}
\end{figure}

We performed an additional set of 5 radiative transfer simulations on the 1 Gyr data varying a single fundamental parameter at a time to explore its effect on the inferred \lya properties of the galaxy. In particular, the dust or neutral gas content were scaled up or down by a factor of 2. In an independent run, the SGS model was switched off to ignore the effects of small-scale turbulence on the radiative transfer. We show only the results for the EW distribution in Fig. \ref{fig:2_variants}. For comparison, the black line shows the fiducial case. The SGS model affects the EW distribution least. Since disabling it effectively makes the Voigt profile narrower and higher, the simulation without SGS model (stars) shows slightly lower EWs.  Photons in the line center are scattered more effectively, and excursions to frequencies far away from the line center become less probable since the velocities of the scattering atoms are statistically lower.

Interestingly, reducing/enhancing the dust content by a factor of 2 (dashed/dashed-dotted line) has the largest effect on the EW distribution, reducing/enhancing the maximum by a factor of a few. Reducing/enhancing the neutral hydrogen content by a factor of 2 (crosses/dashed line) affects the EW distribution only by a factor of 2. The relative sensitivity to dust can be made plausible by considering the fact that to first order, the hydrogen density increases the mean pathlength. Larger pathlengths make absorption exponentially more probable, and the same holds for higher dust densities. On the other hand, the mean pathlength is not necessarily a linear function of the hydrogen density. \cite{Adams1975} has shown that at least for simple geometries, the pathlength scales sub-linearly with the hydrogen density.

\subsection{Dust-free Case}
\begin{figure}
   \centering
   \includegraphics[width=\linewidth]{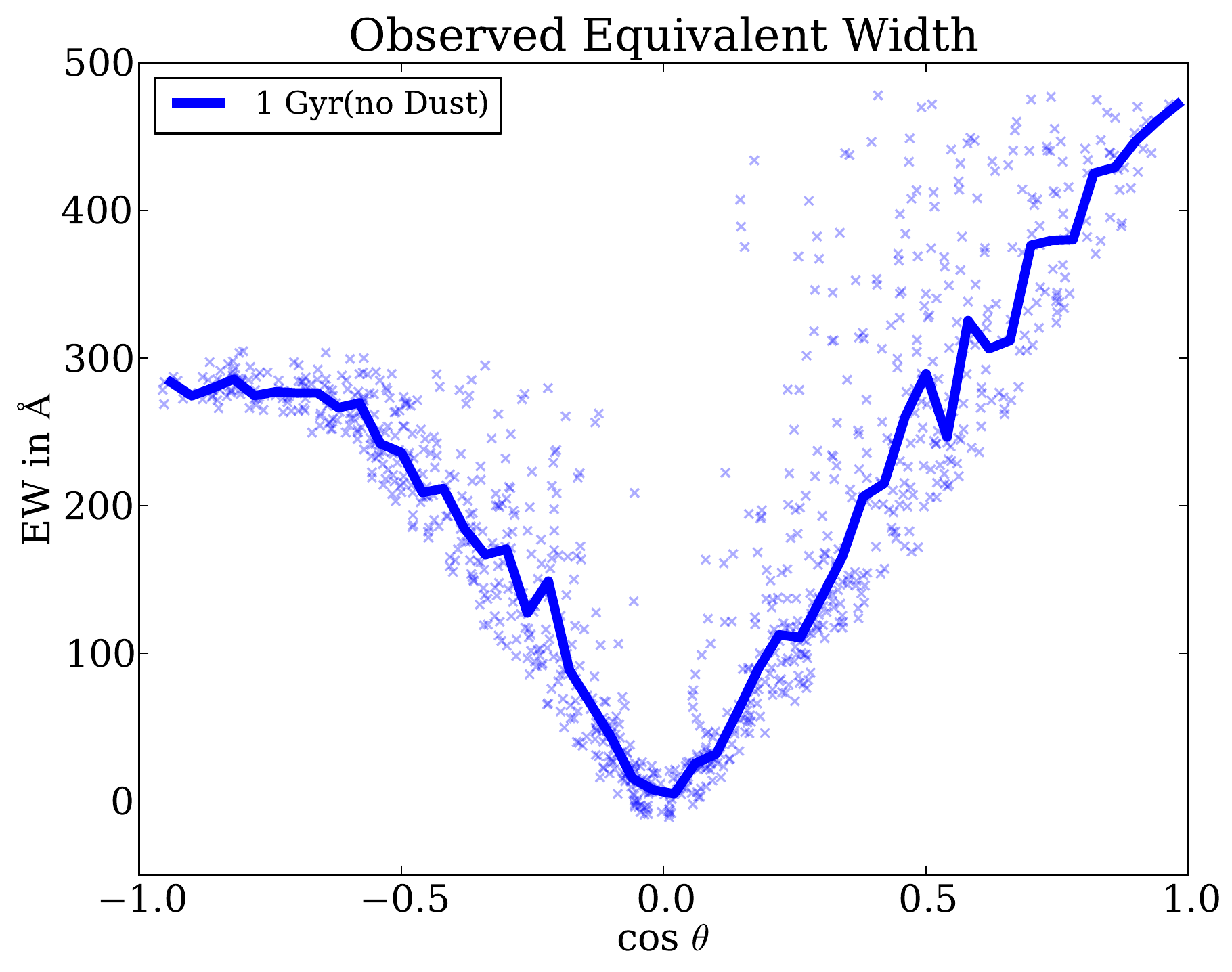}
   
   \caption{Same as Fig. \ref{fig:2}, but for the 1 Gyr snapshot without dust.}\label{fig:2_nodust}
\end{figure}

As a last variation, a simulation ignoring dust in the radiative transfer was performed. This is physically inconsistent with the evolution and history of our simulated galaxy. However, it provides some information about the inclination dependence in extremely metal-poor regimes like e.g. in young starburst galaxies. We show the EW distribution for the 1 Gyr snapshot ignoring dust effects in Fig. \ref{fig:2_nodust}. The absence of dust implies that all photons launched escape the galaxy eventually. While the average EW recovers the chosen intrinsic EW of 200 $\AA{}$, there are large variations between individual lines of sight. A pronounced inclination effect is present here as well. The intrinsic EW is  exceeded by a factor of 2 and more for face-on observers, while edge-on observers see EWs as low as -10 $\AA{}$. The excess above the intrinsic EW in face-on directions is expected. This is a geometrical effect arising from the higher probability for photons to escape in face-on directions compared to edge-
on directions.

\section{Discussion \& Conclusions}
\label{discussion}
We investigated \lya radiative transfer in simulations of an isolated disk galaxy with detailed star formation, feedback and a multiphase ISM, using snapshots of the 'ref' simulation from BSN14 at different timesteps. 

Our results are in broad agreement with VDB12. We find a strong inclination dependence of \lya properties in terms of flux, EW distribution, and spectra. This is readily explained by the different optical depth in directions parallel and perpendicular to the disk. 

As a new result, we find variations of these properties between the snapshots at 1, 1.5, and 2 Gyr. For example, the minimum (maximum) of the observed EW varies from -20 to 0 (70 to 240) \AA{} in our fiducial simulations depending on the snapshot. While this can be partly attributed to the lower neutral hydrogen density of the disk at later times, detailed analysis reveals the importance of supernovae-blown cavities within the disk which reduce the optical depth for photons locally by orders of magnitude (see Fig. \ref{fig:slices}). The transmitting cavities have a lifetime of several 10 Myr corresponding to the period of star formation and stellar feedback. We conjure that the origin of the variations is partly working on timescales of the star-formation activity. The occurrence of cavities is directly connected to the recent local star formation history, and the majority of those that become important as preferred pathways are connected to massive stellar clusters that iteratively host star formation. This 
suggests 
that it is difficult to quantify the statistical inclination dependence of \lya photons for such simulations in a meaningful way, because it would be necessary to derive it not only for a large set of realizations of galaxies, but also to keep track of the evolution of particular emission spots. Compared to our simplified models presented in \cite{Behrens2014} which also feature cavities, the isolated disk galaxy has \lya properties arising from the superposition of multiple, transmitting cavities and a diffuse component, with the cavity component becoming more important at higher dust content. 

On the observational side, our results suggest that the scatter of observed \lya properties can be severely enhanced by this variability on relatively short timescales. To overcome this, one would have to identify individual cavities leaking \lya emission directly, which is difficult at high redshifts where LAEs are not spatially resolved. As described in \citep{Behrens2014}, it might be possible to identify LAEs for which the \lya flux is dominated by cavities by their excess flux at line center (see also Fig. \ref{fig:4}). Statistically, the distribution of EWs in a homogeneous sample of star-forming galaxies identified by an unbiased star formation indicator should show a scatter attributed to the suggested temporal variation.

It will be a further challenge to add two important aspects of realistic galaxies and \lya transport: the immediate surroundings of the galaxy as a source for infalling material, and the IGM scattering \lya photons out of a specific line of sight and frequency. For example, if streams of cool, metal-poor gas were present, we would assume to have a bluer spectrum for photons emerging from the locations where the streams penetrate the disk. Additionally, in these dense streams, the turbulent motion of the gas might play a significant role. Including these features might reduce the variability of the disk, since there might be less lines of sight that have low neutral hydrogen column density. Additionally, following the evolution of the \lya properties over of a galaxy in much smaller timesteps will be necessary to further investigate this temporal variation.

We stress that the inclination dependency of \lya properties is not only found in high-resolution simulations with detailed ISM physics, but also in cosmological simulations with kpc resolution. For example, \cite{Behrens2013} find a $\sim 15$\% increase in flux face-on compared to flux escaping edge-on. This underlines how important it is to understand correlations between inclination and large-scale structure, i.e. galaxy alignment \citep{Hirata2009} and the large-scale density and velocity field \citep{Zheng10,Zheng11}.

{\bf Acknowledgments} C. Behrens and H. Braun were financially supported by the CRC 963 of the German Research Council. The authors thank J. Niemeyer for helpful comments and discussions.

\begin{appendix}
 
\end{appendix}

\end{document}